\documentstyle[psfig,manuscript,aps]{revtex}
\tightenlines
\begin{document}

\title{{\it Ab initio}\ electronic structure calculation of correlated systems:
EMTO-DMFT approach }

\author{L. Chioncel$^1$, L. Vitos$^{2,3}$, I. A. Abrikosov$^4$, J. 
Koll\'ar$^2$, M. I. Katsnelson$^{1,4,5}$, and A. I. Lichtenstein$^1$}

\address{$^1$University of Nijmegen, NL-6525 ED Nijmegen, The Netherlands \\
$^2$ Research Institute for Solid State Physics and Optics, P.O.Box 49,
H-1525 Budapest, Hungary\\
$^3$Applied Materials Physics, Department of Materials Science and
Engineering, Royal Institute of Technology,
SE-10044 Stockholm, Sweden \\
$^4$Uppsala University, P.O.Box 530, S-751 21 Uppsala, Sweden \\
$^5$Institute of Metal Physics, 620219, Ekaterinburg, Russia}
\date{10 March 2003}
\maketitle

\begin{abstract}
We propose a self-consistent method for electronic structure calculations of
correlated systems that combines the local density approximation (LSDA) and
the dynamical mean field theory (DMFT). The LSDA part is based on the exact
muffin-tin orbitals (EMTO) approach, meanwhile the DMFT uses a perturbation
scheme that includes the $T$-matrix with fluctuation exchange (FLEX)
approximation. The current LSDA+DMFT implementation fulfills both 
self-energy and charge self-consistency requirements. We present results on
the electronic structure calculations for bulk $3d$ transition metals (Cr, Fe
and Ni) and for Fe/Cr magnetic multilayers. The latter demonstrates the
importance of the correlation effects for the properties of magnetic
heterostructures.
\end{abstract}


\section{Introduction}

In the {\it ab initio} description of the electronic properties of materials
the most widely used methods are based on the density functional theory (DFT) 
\cite{hohenberg64} implemented within the {\it local spin density approximation}
(LSDA) \cite{kohn65,hedin72} to the exchange and correlation energy. Ground state 
properties of the most of metals, semiconductors, ionic compounds, etc., are 
quantitatively well described by the DFT-LSDA approach. Attempts to apply these 
first principles methods to strongly correlated systems, however, encountered many 
fundamental difficulties \cite{anisimov97a,anisimov97b,lichtenstein98}. Even for 
elemental transition metals, such as Mn, Fe, or Ni, the impact of the correlation 
effects on the electronic structure turns out to be essential \cite{lichtenstein01}. 
Therefore, one of the most challenging problems in the physics of transition metals, 
their alloys and compounds is to develop simple and efficient electronic structure 
methods that go beyond the LSDA by including important many-body effects.

It has proved a fruitful approach to combine the simple Hubbard model with the 
LSDA technique, providing a DFT scheme ''beyond LSDA'' \cite{anisimov97a,anisimov97b,%
lichtenstein98,lichtenstein01}. Unfortunately, the simplest realization of such an approach, 
the LSDA+U scheme \cite{anisimov97a}, cannot describe the many-body effects beyond the 
Hartree-Fock approximation. These effects are connected with the frequency dependence of the
electron self-energy. In order to include dynamical effects the LSDA+U scheme was 
combined with the {\it dynamical mean field theory} (DMFT) \cite{anisimov97b,lichtenstein98}. 
The DMFT maps lattice models onto quantum impurity models subject to a self-consistent 
condition in such a way that many-body problem for crystal splits into one-body 
impurity problem for the crystal and many-body problem for an effective atom. 
In fact, the DMFT, due to numerical and analytical techniques developed to solve 
the effective impurity problem \cite{georges96}, is a very efficient and extensively 
used approximation for energy dependent self-energy $\Sigma (\omega)$. The emerged 
LSDA+DMFT method can be used for calculating a large number of systems with different 
strength of the electronic correlations \cite{lichtenstein01,held02,kotliar01}.
To underline the importance of complete LSDA+DMFT self-consistency we mention that the first 
successful attempt to combine the DMFT with LSDA charge self-consistency 
gave an important insight into a long-standing problem of  phase diagram and localization
in f-electron systems \cite{savrasov01}.

To incorporate the dynamical mean field approach in the band structure
calculation we adopt the {\it exact muffin-tin orbitals} (EMTO) density functional
method. The EMTO theory can be considered as a screened Korringa-Kohn-Rostoker (KKR) 
muffin-tin method, where large overlapping potential spheres are used for accurate
representation of the LSDA one-electron potential. A comprehensive description of 
the EMTO theory and its implementation within the LSDA may be found in Refs.\ 
\cite{andersen94} and \cite{vitos00,vitos01}, respectively.

The paper is organized as follows: Section \ref{forprob} presents a general
formulation of the combined multiple scattering and dynamical mean field
approach. The calculation scheme from Section \ref{calcscheme} illustrates the
multiple scattering solution of the LSDA problem via the EMTO method and the
many-body solution of the DMFT problem via the T-matrix FLEX approach.
First-principles results obtained from EMTO-DMFT calculations are discussed in
Section \ref{results}. The paper is summarized in Section \ref
{conclusion}.

\section{Formulation of the problem}\label{forprob}

The density functional theory reformulates the $N$ electron problem into one
electron problem by considering a non-interacting system, where each electron 
feels an effective potential $v^{\sigma}_{eff}({\bf r})$ created by the rest 
of the electrons and external fields. Thus, within the DFT the solution of 
the original inhomogeneous system is constructed from the one-electron 
Kohn-Sham equations \cite{kohn65}

\begin{equation}  \label{KSeq}
\left[ -\nabla^2\;+\;v^{\sigma}_{eff}({\bf r})\right]\;\Psi^{\sigma}
(\epsilon,{\bf r})\;=\;\epsilon\;\Psi^{\sigma}(\epsilon,{\bf r}),
\end{equation}
where $\sigma$ stands for spin. The many-body part of the effective potential 
$\mu_{xc}^{\sigma}({\bf r})$ is an unknown functional of the spin densities 
$n^{\sigma}({\bf r})=\sum_{\epsilon}|\Psi^{\sigma}(\epsilon,{\bf r})|^2$. The most 
commonly adopted approach for $\mu_{xc}^{\sigma}({\bf r})$ is the {\it local spin 
density approximation} (LSDA), where the effect of interactions between electrons 
is taken into account by substituting locally the real system by the uniform electron 
gas with the density equal to the actual density at point ${\bf r}$. In this paper we  
will not distinguish between different specific forms of the LSDA.

In order to include the many-body correlation effects beyond the LSDA we substitute 
the Kohn-Sham equation (\ref{KSeq}) by the quasiparticle equation

\begin{eqnarray}  \label{ldaUeq}
& &\left[-\nabla^2\;+\;v^{\sigma}_{eff}({\bf r})\right]\;
\Phi^{\sigma}(\varepsilon,{\bf r})\;+  \nonumber \\
&+&\;\sum_{Rl\;mm^{\prime}}|Rlm\sigma\rangle\;
\Sigma^{\sigma}_{Rlm,Rlm^{\prime}}(\varepsilon)\langle Rlm^{\prime}\sigma|\Phi^{%
\sigma}(\varepsilon)\rangle\;=\; \varepsilon\;\Phi^{\sigma}(\varepsilon,{\bf r}),
\end{eqnarray}
where $R,l$ and $m$ denote the lattice sites, the orbital and the magnetic quantum 
numbers, respectively. $|Rlm\sigma\rangle$ are localized orthonormal basis 
functions, e.g. the partial waves for the correlated $l-$channels. In Eq.\ 
(\ref{ldaUeq}) the correlation effects are treated at the DMFT level, where 
the essential many-body self-energy, $\Sigma^{\sigma}_{Rlm,Rlm'}(\varepsilon)$, 
is a local, energy dependent and multi-orbital potential.

Note that Eqs. (\ref{KSeq}) and (\ref{ldaUeq}) are formulated in terms of wave 
functions. Consequently, the DMFT method has already been implemented in several 
techniques based on wave function formalism, like the linear muffin-tin orbital 
method \cite{anisimov97b,lichtenstein98,savrasov01}. 
At the same time, accurate self-consistent methods for 
solving the local Kohn-Sham equation (\ref{KSeq}) in terms of Green's function 
have been developed within the multiple scattering theory \cite{gyorffy72,%
rickayzen80,faulkner80,weinberger90}. The main aim of the present work is to include 
the many-body correlation effects, approximated by means of the DMFT, in the above 
mentioned multiple scattering approach. We note that for a general non-local energy 
dependent potential the multiple scattering theory offers a solution known as the 
optical potential \cite{taylor83}. However, the optical potential is far to 
complicated to be used in realistic computation. Nevertheless, it is proved that 
the non-local potential could be transformed into an one particle energy-dependent 
operator such that it satisfies a similar one-particle equation with local and 
energy independent potential.

\section{The calculation scheme}\label{calcscheme}

\subsection{The one-electron Green's function}

Within the multiple-scattering formalism, the one-electron Green's function is 
defined for an arbitrary complex energy $z$ as 

\begin{equation} \label{oeGf}
\left[z+\nabla^2_{\bf r}-v{^\sigma}_{eff}({\bf r})\right] G^{\sigma}({\bf r},
{\bf r'},z)\;=\;\delta({\bf r-r'}).
\end{equation}
For most of the applications, e.g. standard KKR or LMTO methods, the LSDA effective 
potential from Eq.\ (\ref{oeGf}) is approximated by spherical muffin-tin (MT) wells
centered at lattice sites $R$. Within a particular basis set, the one-electron Green's 
function is expressed in terms of the so-called scattering path operator, 
$g^{\sigma,{\rm LSDA}}_{RL,R'L'}(z)$, as well as the regular, 
$Z^{\sigma}_{RL}(z,{\bf r}_R)$, and irregular, $J^{\sigma}_{RL}(z,{\bf r}_R) $, 
solutions to the single site scattering problem for the cell potential at lattice 
site $R$, {\it viz.}

\begin{eqnarray}\label{g1}
G^{\sigma,{\rm LSDA}}({\bf r}_R+{\bf R},{\bf r}_{R'}+{\bf R'},z)&=&
\sum_{L,L'}Z^{\sigma}_{RL}(z,{\bf r}_R)g^{\sigma,{\rm LSDA}}_{RL,R'L'}(z)
Z^{\sigma}_{R'L'}(z,{\bf r}_{R'})\;-\nonumber\\
&-&\delta_{RR'}\sum_LJ^{\sigma}_{RL}(z,{\bf r}_R)Z^{\sigma}_{RL}(z,{\bf r}_R),
\end{eqnarray}
where $L\equiv(l,m)$ with $l<l_{max}$ (usually $l_{max}=3$) and ${\bf r}_R\equiv
{\bf r}-{\bf R}$ denotes a point around site $R$. The real space representation for 
the scattering path operator for the muffin-tin potential is given by 

\begin{equation}\label{gsmall}
g^{\sigma,{\rm LSDA}}_{RL,R'L'}(z)\;=\;
[\delta_{RR'}\delta_{LL'}{t^{\sigma}}^{-1}_{RL}(z)-B_{RL,R'L'}(z)]^{-1},
\end{equation}
where $t^{\sigma}_{RL}(z)$ stands for the single scattering $t$-matrix and 
$B_{RL,R'L'}(z)$ are the elements of the so-called structure constant matrix.

Unfortunately, the MT based KKR or LMTO methods have limited accuracy. The former 
method uses non-overlapping spherical muffin-tin potentials and constant potential 
in the interstitial, while the latter method approximates the system with overlapping
atomic sphere and neglects completely the interstitial and the overlap between 
individual spheres. Recent progress in the field of {\it muffin-tin orbitals}
theory \cite{andersen94} shows that the best possible representation of the full 
potential in terms of spherical wells may be obtained by using large overlapping 
muffin-tin wells with exactly treated overlaps. Within this so-called {\it exact 
muffin-tin orbitals} method \cite{andersen94}, the scattering path operator is 
calculated as the inverse of the kink matrix defined by

\begin{equation}\label{g2}
K^{\sigma}_{RL,R'L'}(z) \equiv 
\delta_{RR'}\delta_{LL'}D^{\sigma}_{RL}(z) - {\em S}_{RL,R'L'}(z),
\end{equation}
where $D^{\sigma}_{RL}(z)$ denotes the EMTO logarithmic derivative function
\cite{vitos00,vitos01}, and ${\em S}_{RL,R'L'}(z)$ is the slope matrix \cite{andersen94}.
 
Since the energy derivative of the kink matrix, ${\dot K}^{\sigma}_{RL,R'L'}(z)$, 
gives the overlap matrix for the EMTO basis set \cite{andersen94}, the matrix 
elements of the properly normalized LSDA Green's function become \cite{vitos00,vitos01}

\begin{equation}\label{green}
G^{\sigma,{\rm LSDA}}_{RL,R'L'}(z)\;=\;
\sum_{R''L''}g^{\sigma,{\rm LSDA}}_{RL,R''L''}(z){\dot K}^{\sigma}_{R''L'',R'L'}(z) -
\delta_{RR'}\delta_{LL'} I^{\sigma}_{RL}(z),
\end{equation}
where $I^{\sigma}_{RL}(z)$ accounts for the unphysical poles of 
${\dot K}^{\sigma}_{RL,R'L'}(z)$. In the case
of translation invariance Eqs.\ (\ref{g2}) and (\ref{green}) can be transformed to the 
reciprocal space, so that the lattice index $R$ runs over the atoms in the primitive 
cell only, and the slope matrix, the kink matrix, and the path operator depend 
on the Bloch wave vector ${\bf k}$. In this case the total number of states at the 
Fermi level $E_F$ is obtained as

\begin{equation}\label{nos}
N(E_F) = \frac{1}{2\pi i}\sum_{RL,R'L'}\oint
\int_{BZ}G^{\sigma,{\rm LSDA}}_{RL,R'L'}({\bf k},z)\;d{\bf k}\;dz,
\end{equation}
where the energy integral includes the Fermi distribution. The {\bf k}-integral
is performed over the first Brillouin zone, while the $z$-integral is carried out
on a complex contour that cuts the real axis below the bottom of the valence
band and at $E_F$.

\subsection{DMFT Green's function and effective medium Green's function}

To incorporate the many-body effects into the Green's function technique we 
start with the LSDA Green's function matrix (\ref{green}) expressed on the EMTO 
basis set. The LSDA+DMFT Green's function, $G_{RL,R'L'}^{\sigma}({\bf k},z)$,
defined for Bloch vector ${\bf k}$ and energy $z$, is connected to 
the one-electron LSDA Green's function trough the Dyson equation
 
\begin{equation}  \label{G0}
\left[G_{RL,R'L'}^{\sigma}({\bf k},z)\right]^{-1} = 
\left[G_{RL,R'L'}^{\sigma,{\rm LSDA}}({\bf k},z)\right]^{-1} -
\delta_{RR'}{\tilde \Sigma}_{RL,RL'}^{\sigma}(z).
\end{equation}
The local self-energy ${\tilde \Sigma}_{RL,RL'}^{\sigma}(z)$ in Eq.\ (\ref{G0}) 
depends on the so-called effective medium or {\it bath} Green's function 
${\cal G }_{RL,R'L'}^{\sigma }(z)$. This, in turn, is calculated from the 
${\bf k}-$integrated LSDA+DMFT Green's function, $G_{RL,R'L'}^{\sigma }(z)=
\int_{BZ}G_{RL,R'L'}^{\sigma}({\bf k},z)d{\bf k}$, as

\begin{equation} \label{Gtilde}
\left[{\cal G}_{RL,R'L'}^{\sigma }(z)\right]^{-1} =
\left[G_{RL,R'L'}^{\sigma }(z)\right]^{-1}+
\delta_{RR'}{\tilde \Sigma}_{RL,RL'}^{\sigma}(z).
\end{equation}
To find the local self-energy we use a {\it spin polarized} $T$-matrix plus {\it 
fluctuation exchange} (SPTF) approximation \cite{katsnelson01}. The many body
problem is solved on the Matsubara contour, defined by the fermionic frequencies 
$\omega _{n}=(2n+1)\pi T$, where $n=0,\pm1,...$, and $T$ is the temperature. A 
Pade analytical continuation \cite{vildberg77} is used to map between the
complex energies $z$, used in the EMTO iterations, and the complex energies
$i\omega _{n}$, corresponding to the Matsubara frequencies and expressed relative
to the Fermi level $E_F$. Next we describe the solution of the effective impurity 
problem. 

\subsection{The solution of effective impurity problem}

The many-body problem is solved using the SPTF method proposed in Ref. \cite
{katsnelson01}, which is a development of the earlier approach \cite
{katsnelson99}. The SPTF approximation is a multiband spin-polarized
generalization of the fluctuation exchange approximation (FLEX) of Bickers
and Scalapino, but with a different treatment of particle-hole (PH) and
particle-particle (PP) channels. Particle-particle (PP) channel is 
described by a $T$-matrix approach \cite{galitski63} giving a renormalization of
the effective interaction. This effective interaction is used explicitly in the
particle-hole channel. Justifications, further developments and details of
this scheme can be found in Ref.\ \cite{katsnelson01}. Here we present the
final expressions for the electron self-energy. The sum over the ladder
graphs leads to the replacement of the bare electron-electron interaction by
the $T$-matrix which obeys the integral equation

\begin{equation}\label{tmat}
<13|T^{\sigma \sigma ^{\prime }}(i\Omega)|24>=<13|v|24>-T\sum_{\omega
}\sum_{5678}{\cal G}_{56}^{i\sigma }(i\omega ){\cal G}_{78}^{\sigma ^{\prime }}
(i\Omega -i\omega )<68|T^{\sigma \sigma ^{\prime }}(i\Omega)|24>,
\end{equation}
where the matrix elements of the screened Coulomb interaction, 
$<13|v|24>$, are expresed using the average Coulomb and exchange
energies $U,J$ \cite{katsnelson01}.  
In this section, for sake of simplicity, we use the short notation $1=Rlm$. In the 
following we write the perturbation expansion for the interaction (\ref{tmat}). 
The two contributions to the self-energy are obtained by replacing of the bare
interaction by a $T$-matrix in the Hartree and Fock terms

\begin{eqnarray}\label{thtf}
\Sigma_{12}^{\sigma,{\rm TH}}(i\omega )&=&T\sum_{\Omega}\sum_{34\sigma^{\prime}}
<13|T^{\sigma\sigma^{\prime}}(i\Omega)|24>{\cal G}_{43}^{\sigma^{\prime}}
(i\Omega-i\omega) \nonumber\\
\Sigma _{12}^{\sigma,{\rm TF}}(i\omega )&=&-T\sum_{\Omega}\sum_{34\sigma^{\prime}}
<14|T^{\sigma\sigma^{\prime}}(i\Omega)|32>{\cal G}_{34}^{i\sigma^{\prime}}
(i\Omega-i\omega).
\end{eqnarray}
The four matrix elements of the bare longitudinal susceptibility
represents the density-density $(dd)$, density-magnetic $(dm^{0})$,
magnetic-density $(m^{0}d)$ and magnetic-magnetic channels $(m^{0}m^{0})$.
The matrix elements couples longitudinal magnetic fluctuation with density
magnetic fluctuation. In this case the particle-hole contribution to the
self-energy is written in the Fourier transform form 

\begin{equation}\label{selfph}
\Sigma _{12}^{\sigma,{\rm PH}}(\tau )=\sum_{34\sigma ^{\prime }}W_{1342}^{\sigma
\sigma ^{\prime }}(\tau ){\cal G}_{34}^{\sigma ^{\prime }}(\tau ),
\end{equation}
$\tau$ being the imaginary time. The particle-hole fluctuation potential matrix 
$W^{\sigma \sigma ^{\prime }}(i\omega )$ is defined in FLEX approximation 
\cite{bicker89,katsnelson99}

\begin{equation}\label{W}
W^{\sigma \sigma ^{\prime }}(i\omega )=\left( 
\begin{array}{cc}
{W}_{\uparrow \uparrow } & {W}_{\uparrow \downarrow } \\ 
{W}_{\downarrow \uparrow } & {W}_{\downarrow \downarrow }
\end{array}
\right).
\end{equation}
We emphasize that all of the above expressions for the self-energy, in 
the spirit of the DMFT approach, involve the Weiss (bath) Green function 
(\ref{Gtilde}). The total self-energy is obtained from Eqs.\ (\ref{thtf}) 
and (\ref{selfph})

\begin{equation}
\Sigma^{\sigma}(i\omega)=\Sigma^{\sigma,{\rm TH}}(i\omega)+
\Sigma^{\sigma,{\rm TF}}(i\omega)+\Sigma^{\sigma,{\rm PH}}(i\omega).
\end{equation}
Since the LSDA Green's function already contains the average electron-electron 
interaction, in Eqs.\ (\ref{G0}) and (\ref{Gtilde}) the static part of the 
self-energy $\Sigma^{\sigma}(0)$, is not included, i.e. we have

\begin{equation}\label{tildeS}
{\tilde\Sigma}^{\sigma}(i\omega)=\Sigma^{\sigma}(i\omega)-\Sigma^{\sigma}(0).
\end{equation}

\subsection{The charge self-consistency loop within the EMTO-DMFT scheme}

After the self-energy ${\tilde \Sigma}^{\sigma}_{RL,RL'}(z)$ is determined as
the self-consistent solution of the effective impurity problem, the many-body 
LSDA+DMFT Green's function, $G^{\sigma}_{RL,R'L'}({\bf k},z)$, is calculated 
using Eq.\ (\ref{G0}). The EMTO-DMFT number of states at the Fermi level is 
given by the multi center expression ({\ref{nos}), written for the LSDA+DMFT 
Green's function.

The charge and spin densities in the EMTO formalism are represented in one center 
form around each lattice site $R$, i.e.

\begin{equation}
n^{\sigma}({\bf r}) = \sum_{RL} n^{\sigma}_{RL}(r_R) Y_L (r_R), 
\end{equation}
where $Y_L (r_R)$ are the real harmonics. Inside the Wigner-Seitz cell the partial 
components $n^{\sigma}_{RL}(r_R)$ are expressed in terms of the density matrix 
${\mathcal D}^{\sigma}_{R L'L}(z)$ as

\begin{equation}\label{ch}
n^{\sigma}_{RL}(r_R) = \frac{1}{2\pi i} \oint_{E_F}
\sum_{L''L'}C_{L''L'}^LZ^{\sigma}_{Rl''}(z,r_R)\;
{\mathcal D}^{\sigma}_{R L''L'}(z)\;Z^{\sigma}_{Rl'}(z,r_R)dz,
\end{equation}
where $C_{L''L'}^L$ are the real harmonic Gaunt coefficients. The density 
matrix is obtained from the path operator as described in, e.g., Ref.\ \cite{vitos01}.
Within the present EMTO-DMFT scheme the LSDA+DMFT path operator, 
$g_{RL,R'L'}^{\sigma }({\bf k},z)$, is determined according to Eq.\ (\ref{green}) 
using the LSDA+DMFT Green's function and the LSDA overlap matrix. Here we implicitly 
make the assumption that the LSDA+DMFT Green's function can be expanded on the 
same basis set as the one-electron Green's function. In other words, instead of 
the solutions of Eq.\ (\ref{ldaUeq}) for a single scatterer, we use the LSDA 
single-site solutions of Eq.\ (\ref{KSeq}) for a single scatterer to express the 
LSDA+DMFT Green's function in the same form, as we express the one-electron 
Green's function in Eq.\ (\ref{g1}).

Finally, for charge self-consistent calculation we construct the new LSDA effective 
potential from the spin and charge densities, $n({\bf r})=n^{\uparrow}({\bf r})+
n^{\downarrow}({\bf r})$. The Poisson's equation is solved using the spherical 
cell approximation \cite{vitos01}, and the exchange and correlation term is
calculated within the LSDA.

\section{Results and discussion}
\label{results}

The role of the correlation effects in the electronic structure of $3d$
transition metal series is far from being completely understood 
\cite{lichtenstein01}. In general, spin-polarized band structure calculations 
give adequate description of the ferromagnetic ground state for the most of 
metals. At the same time, there are obvious evidences of essentially many-body 
features in photoemission spectra of Fe \cite{katsnelson99}, Co \cite{monastra02}, 
and Ni \cite{lichtenstein01a}. Few examples are the $6$ eV satellite in Ni density
of states, broadening of the angle-resolved photoelectron spectroscopy (ARPES) 
features due to quasiparticle damping, narrowing of the $d$-band, essential 
change of spin polarization near the Fermi level, etc. Although, there are no 
direct experimental information yet, one can assume that the many-body effects 
can also be important in the case of magnetic multilayers and other heterostructures
containing transition metals. The importance of correlation effects on transition 
metal surfaces has already been demonstrated by the STM observation of an {\it 
orbital Kondo resonance} in Cr \cite{kolesnichenko02}. Here we will present 
results obtained using the EMTO-DMFT method in the case of bulk Ni, Fe and Cr 
$3d$ transition metals and for Fe/Cr multilayer structure.

\subsection{Numerical details}

The self-consistent EMTO-DMFT calculations were carried out for the experimental 
ground state crystal structures, i.e. $fcc$ for Ni, and $bcc$ for Fe and Cr. The 
lattice parameters were fixed at the experimental values. The studied Fe/Cr 
multilayer system has $tetragonal$ (001) structure with one type of Fe atoms and two 
different types of Cr atoms. This structural setup describes the situation of 
one Fe layer embedded in few Cr layers. The atoms in the tetragonal unit cell 
were fixed in the ideal positions, and for the lattice parameters we used the 
bulk Cr lattice constant. The LSDA Green's function was calculated for $16$ 
complex energy points distributed exponentially on a semi-circular contour. 
The $k$-point sampling was performed on a uniform grid in the Brillouin zone.
For the LSDA energy functional we used the Perdew-Wang parametrization 
\cite{perdew92} of the results by Ceperley and Alder \cite{ceperley80}. The DMFT 
parameters, average Coulomb interaction $U$, exchange energy $J$ and temperature 
$T$, used in the present calculation are listed in the last three columns of 
Table\ \ref{tab1}. 

From the self-consistent density of states (DOS) we have determined the magnetic
moment $\mu$ and the electronic specific heat coefficient $\gamma$. The latter
is given by relation 

\begin{equation}\label{lamb}
\gamma =\pi ^{2}k_{B}^{2}N(E_F)(1+\lambda )/3,
\end{equation}
where $N(E_F)$ is the electronic DOS at the Fermi level, $(1+\lambda)$
is the mass enhancement factor caused by the electron-phonon interaction. This 
factor in the case of Ni was estimated to be $1.24$ \cite{anderson79}.
The present theoretical results for the self-consistent magnetic moments and
electronic specific heats, along with the available experimental data are 
listed in Table\ \ref{tab1}.

\subsection{Ni}

It has been shown that the main peculiarities of the experimental Ni 
photoemission spectra can be understood within the framework of the LSDA+DMFT 
approach \cite{lichtenstein01a}. An exact quantum Monte-Carlo (QMC) solution 
of the effective impurity problem \cite{lichtenstein01a} gives impressive
quantitative agreement between the experimental and computational data, both
for photoemission spectra and for temperature dependent magnetic properties.
Here we will show that the perturbative SPTF approach, employed in the present 
EMTO-DMFT method, also reproduces the main correlations effects beyond LSDA in 
Ni, e.g. the narrowing of the band, reduction of the exchange splitting and 
the appearance of the $6$ eV satellite.  

The EMTO-DMFT density of states for $fcc$ Ni is shown on Fig.\ \ref{nidos}. 
For the present choise of the average Coulomb interaction $U=3eV$,
the position of the $6$ eV satellite is shifted to the lower energy. 
This shift and the large broadening of the resonance is due to the perturbative approach 
of the solver of the effective impurity problem \cite{katsnelson01}. 

According to the present result, at $T=500$ K, i.e. at $T/T_c \approx 0.8$, 
the reduction of the exchange splitting relative to the LSDA value is $43\%$. 
This is in good agreement with $40\%$ estimated from experimental 
data corresponding to the same temperature. Our results are also in agreement 
with ARPES measurements \cite{eastman78,maetz82}. The EMTO-DMFT magnetic moment
of $0.42\mu_{B}$, see Table\ \ref{tab1}, represents a reduction of $30\%$ from 
the LSDA value. This reduction is comparable with $25\%$ evidenced from experimental 
magnetic moments.

Apart from the many-body self-consistency, the present implementation fulfills 
also the charge self-consistency. This allows us to monitor the effect of DMFT 
on the LSDA charge and magnetic moment densities in the real space. Fig.\
\ref{magm} shows the LSDA and LSDA+DMFT magnetic moment densities in Ni 
along the $\left\langle 110\right\rangle$ direction, which corresponds to the 
nearest-neighbor distance in the $fcc$ unit cell. Comparing the two densities
one can interpret the reduction of the magnetic moment in the LSDA+DMFT approach
as a slight narrowing of the real space extension of the $d$ wave functions.

The energy dependence of the self-energy for Ni, plotted on Fig.\ \ref{nisigm}, 
near the Fermi level shows the typical Fermi liquid behavior. For the imaginary
part we have $-Im\ \Sigma (E)\propto E^{2}$, meanwhile the real part of the 
self-energy has a negative slope $\partial Re\ \Sigma (E)/\partial E<0$, where 
$E$ is the electron energy relative to the Fermi level.

Within the LSDA for the electronic specific heat coefficient we obtain $5.43$ 
mJ/K$^2$mol, which underestimates the experimental value from Ref.\ 
\cite{kittel66} by more than $20\%$. This LSDA value is in good agreement with 
previous calculations \cite{steiner92,bacalis97}. On the other hand, within the 
LSDA+DMFT approach the electronic specific heat coefficient is $6.78$ mJ/K$^2$mol, 
which reduces the discrepancy between theory and experiment by $19\%$.

\subsection{Fe}

In Fig \ref{fedos} we compare the LSDA and the LSDA+DMFT density of states for
$bcc$ Fe. The present LSDA magnetic moment $2.25\mu_{B}$ is reduced to 
$2.23\mu_{B}$ within the LSDA+DMFT approach. We note that our LSDA magnetic
moment is in excellent agreement with the one obtained in former {\it ab initio}
calculation  \cite{singh91}. The electronic specific heat coefficient increase 
from the LSDA value of $2.43$ mJ/K$^2$mol to $2.53$ mJ/K$^2$mol in the 
LSDA+DMFT calculation. These values can be compared with the experimental data 
in the range of $3.11-3.69$ mJ/K$^2$mol \cite{phillips71}. We can see that in
$bcc$ Fe the correlation effects are much less pronounced than in $fcc$ Ni.
This is due to the Fe large spin splitting and the $bcc$-structural dip in the 
density of states \cite{lichtenstein01a}. The energy dependence of the self-energy 
of Fe, see Fig.\ \ref{fesigm}, corresponds again to the Fermi liquid behavior, similar 
to the case of Ni.

The $d$ band exchange splitting in Fe in the LSDA+DMFT is slightly decreased 
in comparison with the LSDA result. The temperature dependence of the exchange 
splitting in Fe was determined by spin-resolved photoemission spectroscopy 
\cite{kisker84}. The experimental results show a very weak temperature dependence
of the exchange splitting in the temperature range $0.3-0.85T_c$, where $T_c=1043$ K.
Our calculations for three different temperatures $T/T_c=0.3, 0.6$ and $0.8$ 
show almost constant $d$ band exchange splitting, in perfect agreement with
experiment \cite{kisker84} and previous DMFT(QMC) calculations \cite{lichtenstein01}.

\subsection{Cr}

The effect of correlations in the case of $bcc$ Cr is manifested through a small 
enhancement of the density of states at the Fermi level, as shown in Fig.\ 
\ref{crdos}. The value of electronic specific heat coefficient is $2.88$ 
mJ/K$^2$mol, which represents an improvement of about $40\%$ relative to the 
LSDA value of $2.07$ mJ/$K^2$mol. Our LSDA+DMFT result still underestimates the
experimental non-magnetic data of $3.5$ mJ/K$^2$mol \cite{fawcett88} by
$15\%$. Although, for the bulk Cr the correlation effects are not very important,
one can expect strong correlation effects at Cr surfaces, in the light of the 
observation of essentially many-body phenomenon, the orbital Kondo resonance
at Cr(001) surface \cite{kolesnichenko02}.

To demonstrate the correlation effect on the real space charge distribution in Fig.\
\ref{chdLSDA-DMFT} we have plotted the difference of the LSDA+DMFT and LSDA charge 
densities in the $bcc$ (110) plan. As one can see the main effect of the DMFT 
charge self-consistency is a redistribution of charge density, suggesting a 
suplimentary accumulation of $d$ electrons due to correlation effects inside 
the muffin-tin spheres and a depletion of density in the interstitial region.  

\subsection{Fe/Cr multilayer}

In order to test the opportunities of the current LSDA+DMFT approximation further, 
we have applied this to magnetic heterostructures of alternating magnetic and
nonmagnetic layers. The most remarkable property of these systems is the giant 
magnetoresistance (GMR) measured for a parallel/antiparallel configuration of 
the magnetic moments belonging to the different layer by application of a magnetic 
field \cite{fert89}. Experimentally was found that the magnetic multilayers grown 
epitaxially shows an enhancement of the electronic contribution to the low 
temperature specific heat \cite{revaz02}. Standard electronic band structure
calculations could not reproduce this enhancement \cite{kulikov97}, which can be 
an evidence of the correlation effects.  Our aim is to check whether the 
correlation effects considered in the EMTO-DMFT approach can lead to an 
essential renormalization of the density of state at the Fermi level $N(E_F)$. 

In the LSDA+DMFT calculations for Fe/Cr multilayer we have chosen 
different values for the average Coulomb interaction and exchange energy for Fe 
and the two types of Cr atoms. These values are listed in Table\ \ref{tab1}. 
Although this choice is motivated by the fact that correlation effects for these
atoms could be similar in this structure, the present results are more qualitative 
than quantitative.

The layer resolved LSDA and LSDA+DMFT densities of states for Fe/Cr multilayer
are plotted in Fig.\ \ref{mldos}. Comparing the LSDA and LSDA+DMFT DOS one can 
see that the correlation effects produce a strong peak at the Fermi level. This
can give a qualitative explanation of the heat capacity data discussed above.

The calculated LSDA and LSDA+DMFT magnetic moments are shown in Fig.\ \ref{momsc}. 
Our LSDA result is in  agreement with previous electronic structure calculations
of Fe impurities in a Cr surrounding \cite{mirbt97}. 
We have found that within the LSDA the magnetic multilayer structure consist of
ferromagnetically coupled Fe layers with the magnetic moments of $1.72\mu_B$ 
per atom. The Cr spacers have very small magnetic moments per atom, $-0.05\mu_B$ and 
$0.09\mu_B$, respectively, and they are oriented antiferromagnetically.
This result is in accordance with previous {\it ab initio} study on Fe/Cr superlattices 
\cite{ounadjela91}. The LSDA Fe magnetic moment is drastically reduced compared with 
its value in bulk, which is attributed to $d-d$ band hybridization between the Fe and 
Cr states.

The results of the LSDA+DMFT calculations are essentially different from 
that in LSDA. In particular, the correlation effects result in a slightly
increase of the Fe layer magnetic moment and a 
significant polarization of the Cr spacers, see Table\ \ref{tab1}.
In the case when the correlation effects are considered for all the atoms
the selfonsistent calculation shows the same trend for the magnetic moments:     
$1.85\mu_B$ per Fe atom, $0.42\mu_B$ per Cr1 and  $-0.19\mu_B$ per Cr2 
respectively.  These results leads us to the conclusion that 
the correlation effects induce a strong polarization on the first Cr layer, which is almost 
non-magnetic according to the LSDA calculations. The appearance of the Cr magnetic moments 
can be attributed to the correlation induced narrowing of the $d$ band, together with the 
Fe-Cr $d-d$ hybridization mechanisms. In the presence of the correlations the Fe-Cr $d-d$ 
hybridization is much stronger than the Cr$_1$-Cr$_2$ $d-d$ hybridization. This can be seen 
from majority spins channel of DOS from Fig.\ \ref{mldos}, where Fe $d$ and Cr$_1$ $d$ have 
pronounced peaks at the Fermi level. Due to this significant change in the spin up
d-chanell of DOS of Fe Fig.\ \ref{mldos}, the spectral weight is transfered close to
Fermi level and the effective exchange interaction between the Fe and the first Cr layer is 
changed becoming ferromagnetic. Further investigation of this magnetic ordering
as function of different parametes $U, T$ is important for the nature of the magnetic
coupling in multilayer systems.  

Several theoretical approaches have been used to explain the magnetic
properties of such superlattice structures. Many of these approaches are
based on the RKKY-like model \cite{yafet87,wang91}, tight-binding models \cite
{stoeffler90} and, recently, on the results of  {\it ab initio} electronic
structure calculations \cite{ounadjela91}. The magnetic coupling studied in the
framework of these models was shown to result form the interplay between the
direct $d-d$ hybridization of Fe and Cr atoms and indirect exchange through
the $sp$ electrons. The $sp-d$ coupling \cite{ounadjela91} was found to be
reminiscent of the RKKY interaction only for superlattices with more than
four Cr layers. 

The calculated electronic specific heat coefficients are listed in Table\ \ref{tab1}.
The present LSDA+DMFT value of $7.84$ mJ/K$^2$mol is in good agreement with
recent experimental study \cite{revaz02}. %Note that the value of $\gamma$

Finally, it is worthwhile to emphasize that the enhanced DOS at the Fermi level,
having a many-body correlation origin, can play an important role in the
GMR, since this DOS enhancement is strongly spin-dependent. It is more
effective for the majority electrons of Fe and Cr$_1$ (Cr$_1$ are the
one closer to the Fe layer), giving the result of a quasiparticle peak
centered at the Fermi level. Our founding is in good agreement with the
tendency of the enhancement of electronic contribution to the
specific heat in Fe/Cr magnetic multilayers \cite{revaz02}.

\section{Conclusions}
\label{conclusion}

In this paper we present a LSDA+DMFT scheme on the Exact Muffin-Tin Orbitals
basis set. The present EMTO formalism allows us to combine the many body problem 
with the standard screened KKR(LSDA) method in a self-consistent manner. The many-body 
self-consistency is reached trough self-consistency of self-energy, meanwhile the 
self-consistent charge density is obtained in the conventional LSDA framework.
The results of our EMTO-DMFT calculation are summarized in Table\ \ref{tab1}, and they
are in good agreement with former LSDA+DMFT implementations using the LMTO basis set.

Correlations effects in multilayer systems are important for transport
properties, giving rise to an enhancement of density of states at the Fermi
level. We have studied a simple magnetic multilayer system, and have shown that 
the effect of correlation is to induce magnetism in non magnetic spacers. We attribute 
this structure and temperature dependent polarization mechanism to simultaneous: (i)
Fe-Cr $d-d$  hybridization and (ii) narrowing of electronic $d$ bands due
to many-body correlation effects. We note that the latter mechanism, narrowing of the
Cr $d$ band, is significantly stronger than the former one.

\section*{Acknowledgments}

L.C. acknowledges the financial supports from: Marie Curie host fellowship 
Contract Number HPMT-CT-2001-00281, the
{\it Computational Magnetoelectronics} RTN project (HPRN-CT-2000-00143) during
his visit in the Research Institute for Solid State Physics and Optics, Budapest,
and Kavli Institute for Theoretical Physics through the National Science Foundation
Grant No. PHY99-07949. M. I. K. acknowledges the financial support from Russian 
Science Support Foundation. 

This work was supported by the Netherlands Organization for Scientific Research 
(NWO), grant NWO 047-008-16. The Swedish Natural Science Research Council, the 
Swedish Foundation for Strategic Research and Royal Swedish Academy of Sciences 
are acknowledged for financial support. Part of this work was supported by the 
research project OTKA T035043 of the Hungarian Scientific Research Fund and the
Hungarian Academy of Science.

\begin{table}[tbp]
\begin{tabular}{c|ccccc|ccc}
& $\mu_{\rm LSDA}$ & $\mu_{\rm DMFT}$ & $\gamma_{\rm LSDA}$ & $\gamma_{\rm DMFT}$ & $ \gamma_{\rm expt.}$ & $T$ & $U$ & $J$ \\
& ($\mu_B$) & ($\mu_B$) & (mJ/K$^2$ mol) & (mJ/K$^2$ mol) & (mJ/K$^2$ mol) & (K) & (eV) & (eV) \\ \hline\hline
Ni        & 0.63 & 0.42 & 5.43 & 6.78 & 7.02$^a$      & 500 & 3 & 0.9 \\
Fe        & 2.25 & 2.23 & 2.43 & 2.61 & 3.11,3.69$^b$ & 325 & 2 & 0.9 \\
Cr        &    - &    - & 2.07 & 2.88 & 3.5$^c$       & 250 & 2 & 0.9 \\
\hline
Fe(Fe/Cr)     &  1.72 &  1.75 &          &          &                   &     & 2 & 0.9 \\
Cr$_1$(Fe/Cr) & -0.05 &  0.15 & 6.90$^*$ & 7.84$^*$ & $8.7\pm0.7^{*,d}$ & 300 & 0 & 0.0 \\
Cr$_2$(Fe/Cr) &  0.09 & -0.11 &          &          &                   &     & 0 & 0.0
\end{tabular}
$^a$ Ref.\ \cite{kittel66} \\
$^b$ Ref.\ \cite{phillips71}             \\
$^c$ Non-magnetic Cr, Ref.\ \cite{fawcett88}. \\
$^d$ Ref.\ \cite{revaz02} \\
$^*$ Values corresponding to the magnetic multilayer.
\caption{Theoretical magnetic moments, $\mu$, and electronic specific heat coeficients,
$\gamma$, calculated at LSDA and LSDA+DMFT levels. For comparison some experimental
electronic specific heat coeficients are also listed. The experimental values 
for $\gamma$ include the enhancement due to electron-phonon coupling, but this enhancement 
is not included for the calculated band structure values for Fe, Cr and Fe/Cr multilayer.
The theoretical $\gamma$ values for Ni are corrected with $1+\lambda=1.24$ according to 
Eq.\ (\ref{lamb}) and Ref.\ {\protect\cite{anderson79}}. In the last three columns the 
parameters used in the self-consistent EMTO-DMFT calculations are listed.}
\label{tab1}
\end{table}

\begin{figure} 
\centerline{\psfig{figure=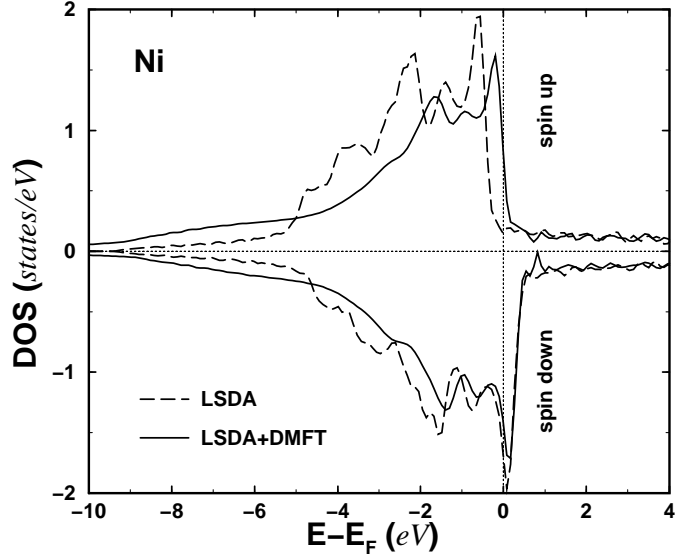,height=3.0in}}
\caption{The LSDA (dashed line) and LSDA+DMFT (solid line) densities of states 
for $fcc$ Ni calculated using the EMTO-DMFT method. A significant reduction of 
the exchange splitting can be evidenced.}\label{nidos}
\end{figure}

\begin{figure}
\centerline{\psfig{figure=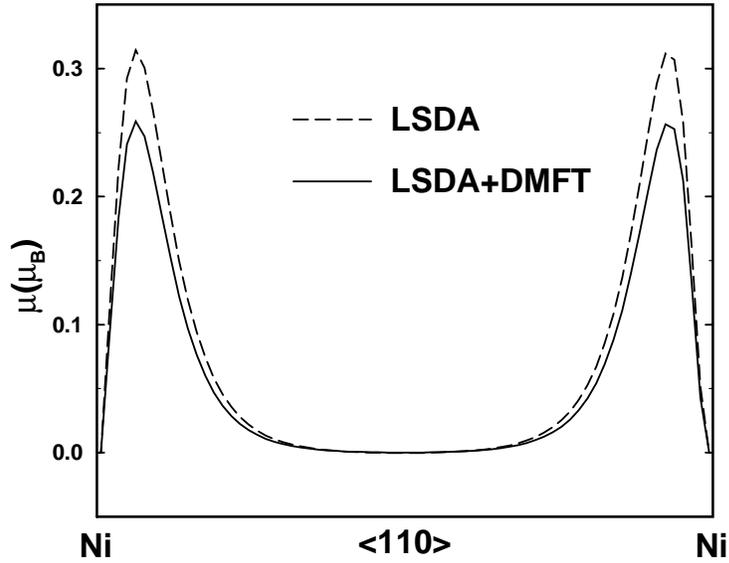,height=3.0in}}
\caption{Magnetic moment densities of Ni along the nearest-neighbor distance 
in $fcc$ lattice, calculated within the LSDA (dashed line) and the LSDA+DMFT
(solid line).} \label{magm}
\end{figure}

\begin{figure}
\centerline{\psfig{figure=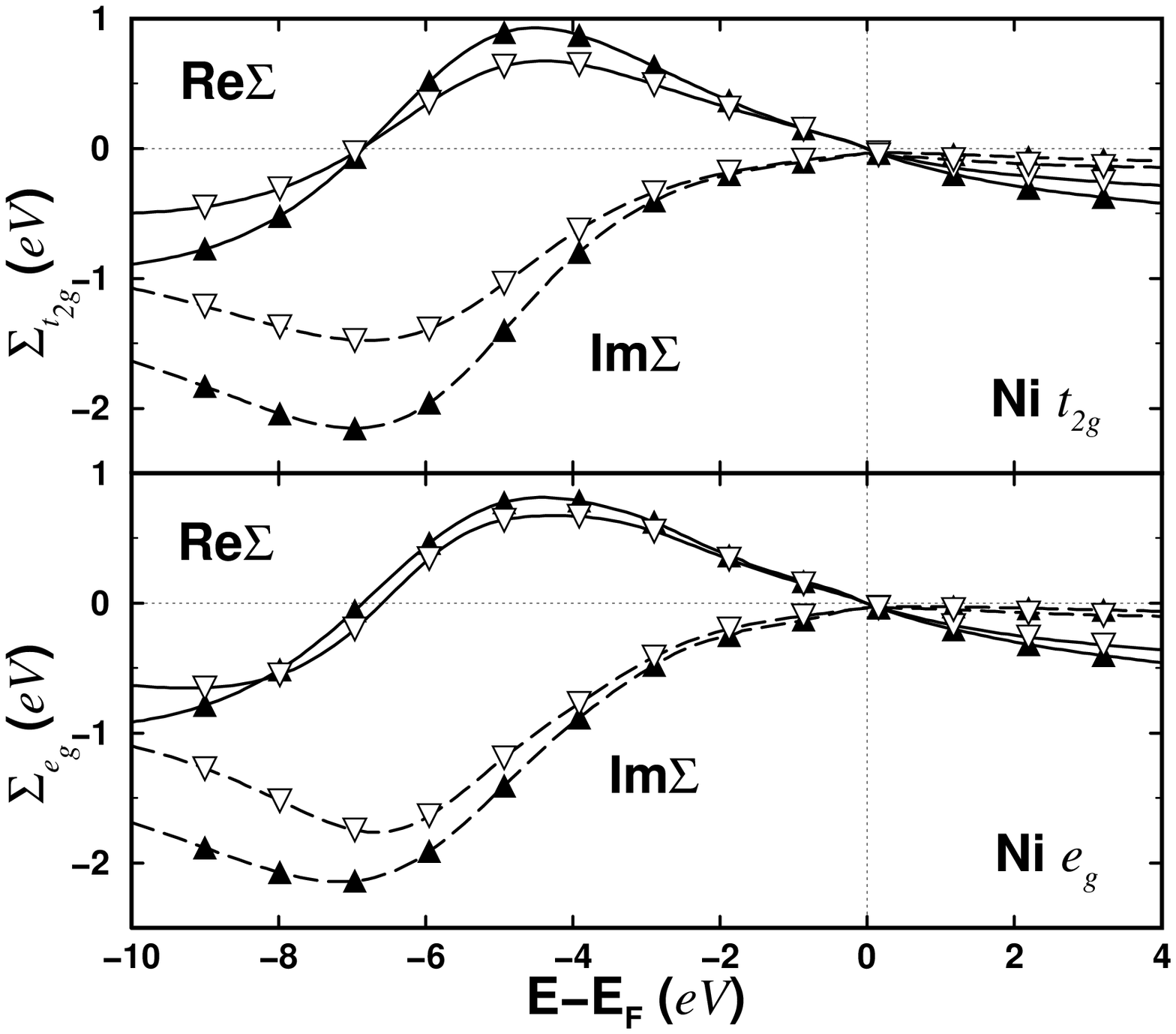,height=3.0in}}
\caption{Spin up (open symbols) and down (closed symbols) self energies for
Ni for $t_{2g}$ (upper panel) and $e_g$ (lower panel) orbitals.} \label{nisigm}
\end{figure}

\begin{figure}
\centerline{\psfig{figure=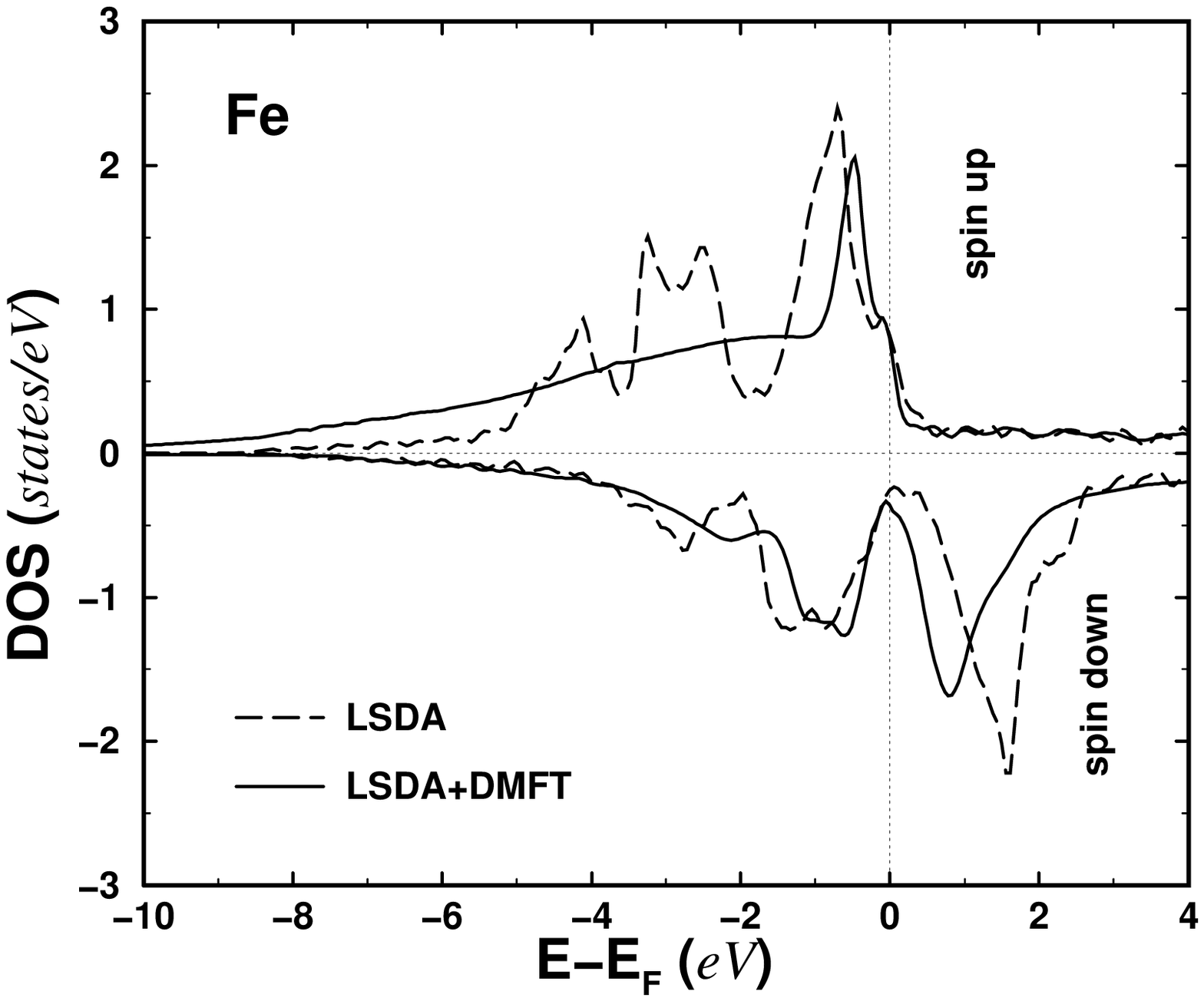,height=3.0in}}
\caption{The LSDA (dashed line) and LSDA+DMFT (solid line) densities of states
for $bcc$ Fe calculated using the EMTO-DMFT method.} \label{fedos}
\end{figure}

\begin{figure}
\centerline{\psfig{figure=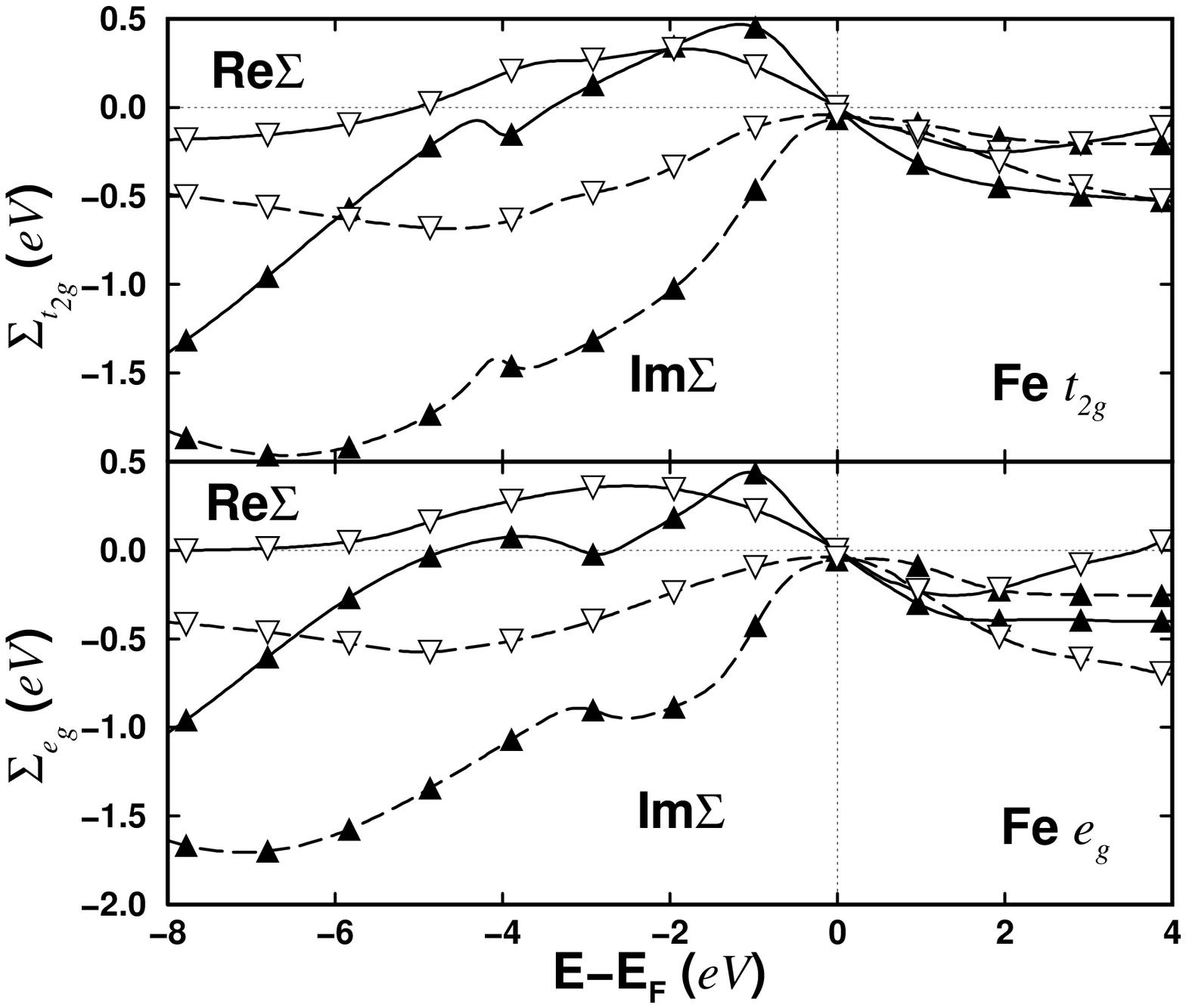,height=3.0in}}
\caption{Spin up (open symbols) and down (closed symbols) self energies for
Fe for $t_{2g}$ (upper panel) and $e_g$ (lower panel) orbitals.} \label{fesigm}
\end{figure}

\begin{figure}
\centerline{\psfig{figure=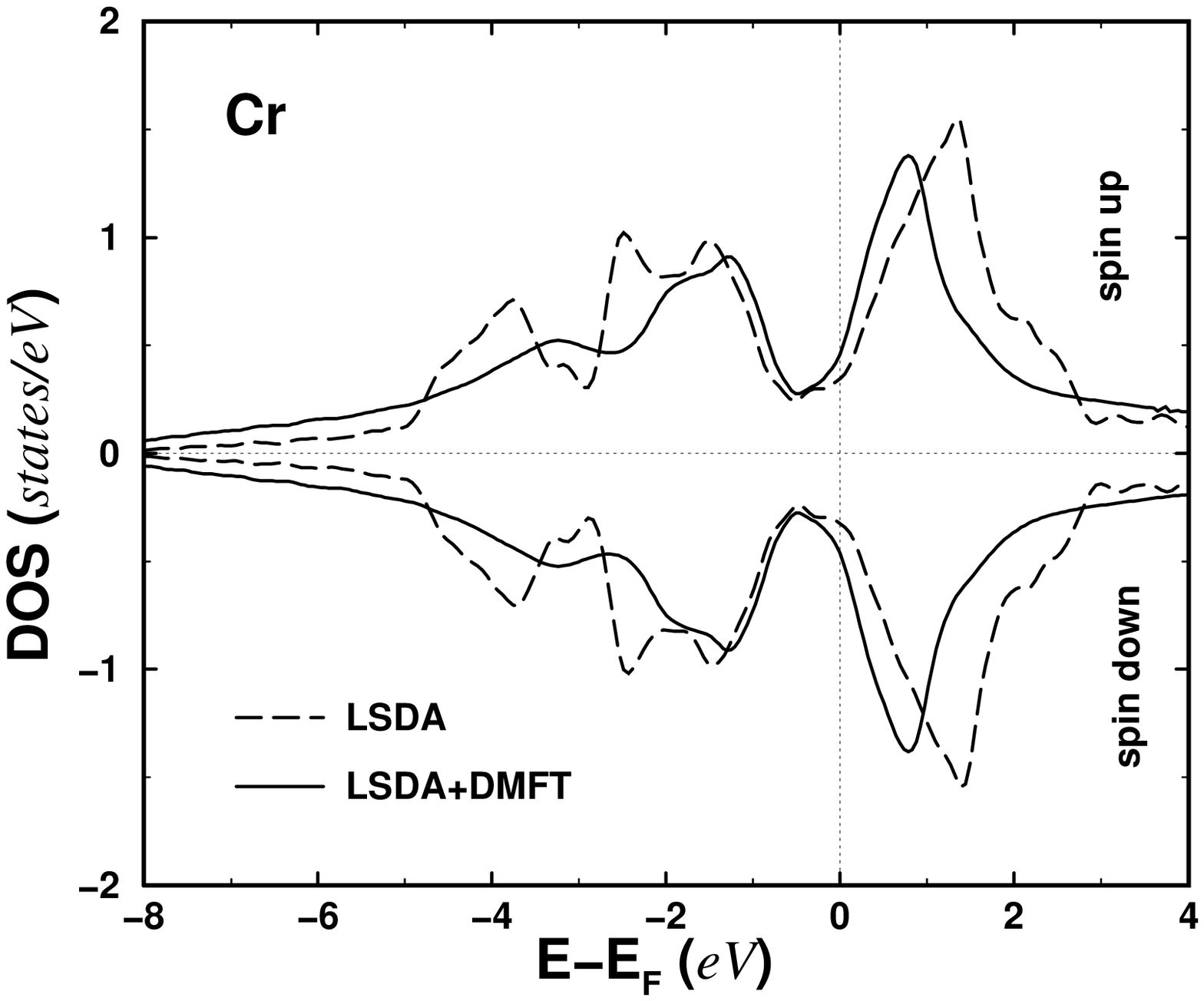,height=3.0in}}
\caption{The LSDA (dashed line) and LSDA+DMFT (solid line) densities of states
for $bcc$ Cr calculated using the EMTO-DMFT method.} \label{crdos}
\end{figure}

\begin{figure}
\centerline{\psfig{figure=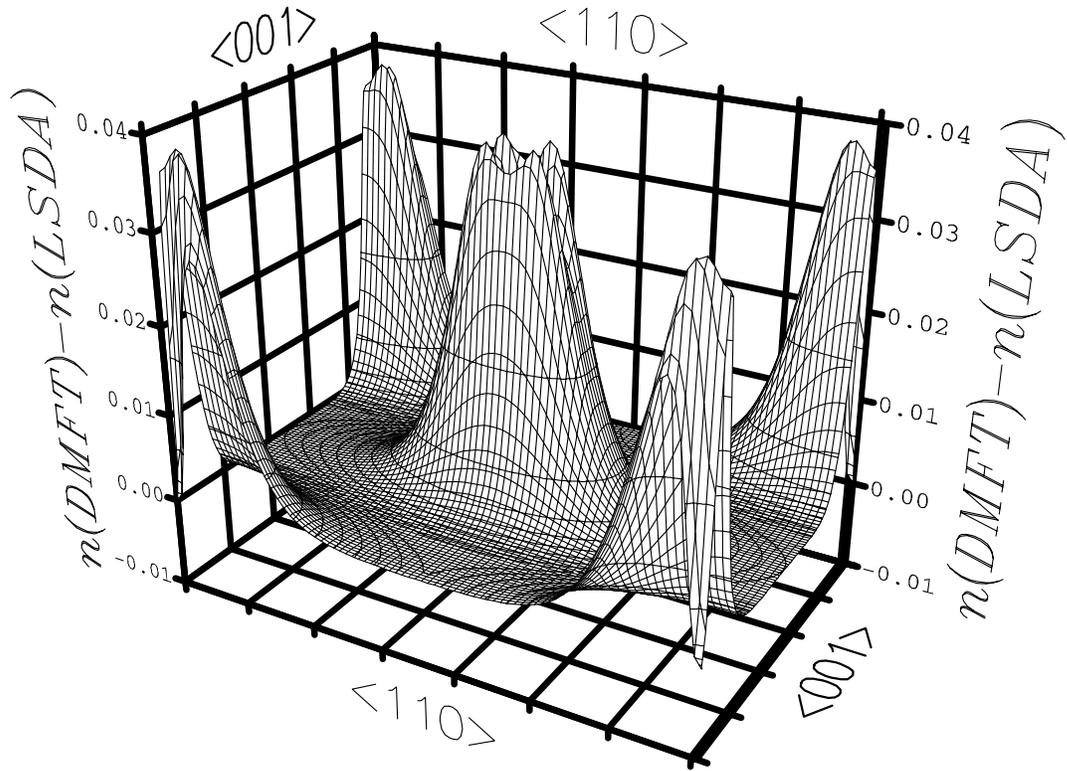,height=4.0in}}
\caption{The effect of DMFT on the LSDA charge density of $bcc$ Cr.}\label{chdLSDA-DMFT}
\end{figure}

\begin{figure}
\centerline{\psfig{figure=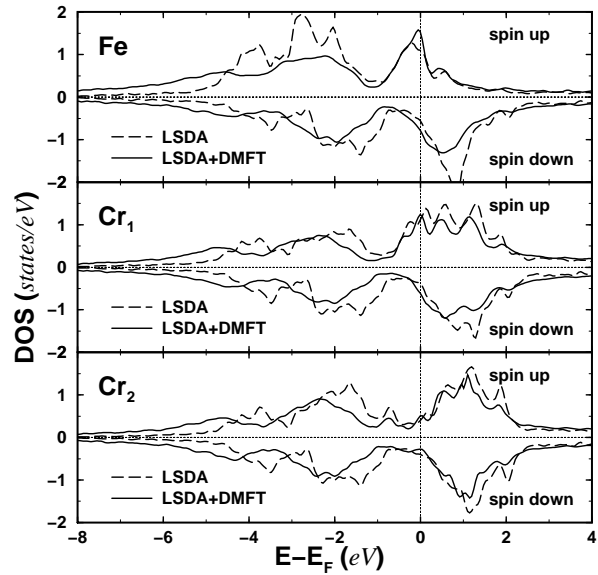,height=3.0in}}
\caption{The layer resolved LSDA (dashed line) and LSDA+DMFT (solid line) densities of 
states for Fe/Cr$_1$/Cr$_2$/Cr$_1$/Fe multilayers.} \label{mldos}
\end{figure}

\begin{figure}
\centerline{\psfig{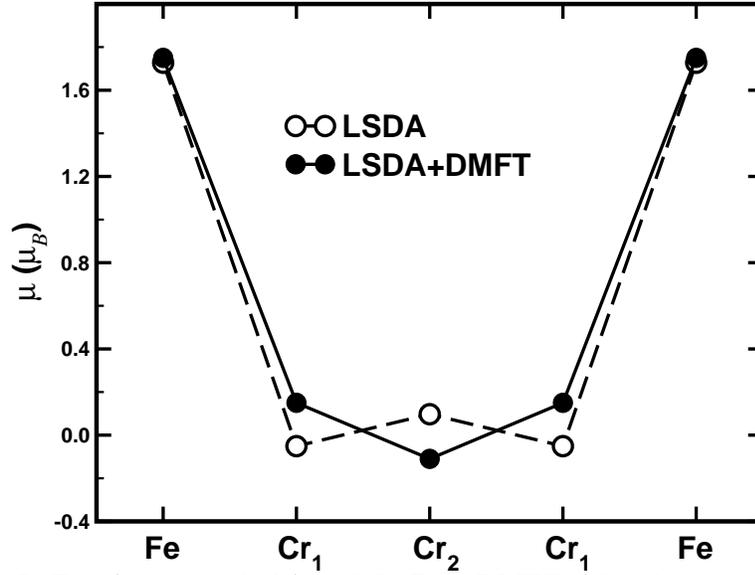}}
\caption{The LSDA (open symbols) and LSDA+DMFT (closed symbols) magnetic moments per atom
for Fe/Cr$_1$/Cr$_2$/Cr$_1$/Fe multilayers.}\label{momsc}

\end{figure}

\end{document}